\documentstyle[12pt]{article}
\title{Low energy parameters of the $K\overline{K}$ and $\pi\pi$
scalar-isoscalar
interactions}
\author{R. Kami\'nski and L. Le\'sniak \\Department of Theoretical
Physics, \\
Henryk Niewodnicza\'nski Institute of Nuclear Physics, \\
PL 31-342 Krak\'ow, Poland}

\begin{document}

\maketitle
\begin{abstract}
Threshold expansions of the $\pi\pi$ and $K\overline{K}$ spin 0 and
isospin 0 scattering
amplitudes are performed. Scattering lengths, effective ranges and
so--called
volume parameters are evaluated. Good agreement with the existing
experimental
data for the $\pi\pi$ scalar--isoscalar amplitude is found.
An importance of future accurate measurements of the
$K\overline{K}$ threshold parameters is stressed.
New data are needed to understand the basic features of the scalar
mesons.
\end{abstract}

Kaon--antikaon interactions are very poorly known.
A characteristic feature of the $K\overline{K}$ interactions is a
presence of the
annihilation processes in which a creation of the $\pi\pi$  pairs
plays a very important role. Thus the $K\overline{K}$ and $\pi\pi$
channels are
coupled together and should be treated simultaneously. Our knowledge
of the meson--meson interactions is based mainly on the reactions in
which the kaon or pion pairs are produced. The production processes
of the scalar mesons $f_{0}(975)$ and $a_{0}(980)$ (which both decay
into the $K\overline{K}$ pairs)
have been studied in many experiments \cite{kon,karch} and new
experiments like those at COSY (J\"ulich) \cite{oelert},
DA$\Phi$NE (Frascati) \cite{franzini,bramon} and CEBAF (Newport News)
are planned.
Unfortunately the existing $K\overline{K}$
and $\pi\pi$ data are not sufficiently precise to construct a unique
model explaining the nature of the poorly known scalar mesons.
Therefore different theoretical approaches to this question exist
(see for example refs. [6--12]).

In order to compare various models of the $K\overline{K}$ interactions
we propose
to calculate in future for each theoretical framework the low energy
$K\overline{K}$
parameters using the effective range approximation known for example
from the studies of the nucleon--nucleon interactions \cite{brown}.
These parameters are crucial in understanding the nature of the
$K$--$\overline{K}$ interactions.
The importance of computing the threshold parameters has been
also recently stressed by T\"ornqvist \cite{trnqv}.
The masses of the $f_{0}(975)$ and $a_{0}(980)$ mesons are very close
to the $K\overline{K}$ threshold.
Therefore these mesons are frequently interpreted as the
quasibound states of the $K\overline{K}$ pairs [15--19].
In ref. \cite{cann92} the $K\overline{K}$ scalar--isoscalar scattering
length has been already calculated using a separable potential formalism.
Then Wycech and Green have used its value to discuss a production of the
kaonic atoms \cite{wycech}.
More recently we have extended the calculations of the scalar--isoscalar
$K\overline{K}$ and $\pi\pi$ scattering amplitudes using the
relativistic approach \cite{klm}.
A simple rank--one separable potential has been used to describe the
$K\overline{K}$
interaction and a rank--two potential in the $\pi\pi$ channel.
Choosing the
rank-two potential responsible for the coupling of two channels we have
obtained very good fits to the data starting from the $\pi\pi$ threshold
up to 1400 MeV, thus fully covering the interesting region of the
$K\overline{K}$ threshold
near 1 GeV \cite{marsylia}. In this procedure we have been able to fix
the parameters of the meson--meson interactions. As a next step
we report the results of the calculations of the threshold parametrs
for the $K\overline{K}$  and $\pi\pi$ interactions in the spin and
isospin zero state.

We use the effective range expansion in the $\pi\pi$ and $K\overline{K}$
channels:

\begin{equation}
k\,cot \delta = \frac{1}{a} + \frac{1}{2} r k^{2} + v k^{4} + O(k^{6}),
\label{kcot}
\end{equation}
where $\delta$ is the scattering phase shift, $k$ is the relative meson
momentum,\\ $a$ is the scattering length, $r$ is the effective range of
the interaction and the parameter $v$ can be related to the shape of the
intermeson potentials.

The low momentum expansion of the phase shift has a polynomial form:

\begin{equation}
\delta = \alpha k + \beta k^{3} + \gamma k^{5} + O(k^{7}).
\label{delta}
\end{equation}
The coefficients $\alpha, \beta, \gamma$ can be obtained from the low
momentum
expansion of the scattering amplitudes calculated in ref. \cite{klm}.

Above the $K\overline{K}$ threshold we define the complex
$K\overline{K}$ phase shift \mbox{$\delta=\delta_{K} +i\rho$},
 where $\delta_{K}$ is the $K\overline{K}$ phase shift and $\rho$ is
related to the inelasticity parameter

\begin{equation}
\eta=e^{-2\rho}.
\label{rho}
\end{equation}
In the $K\overline{K}$ channel the expansions (\ref{kcot}) and
(\ref{delta}) can still be valid
if we make the parameters $a, r, v$ and
$\alpha, \beta, \gamma$ complex. From Eqs. (\ref{kcot}) and
(\ref{delta}) one can derive the
following relations between these parameters:

\begin{equation}
a=\alpha,
\label{a}
\end{equation}
\begin{equation}
r=-\frac{2}{3}\alpha-2\frac{\beta}{\alpha^2},
\label{r}
\end{equation}
\begin{equation}
v=-\frac{1}{45}\alpha^3-\frac{1}{3}\beta+\frac{\beta^2}{\alpha^3}-
\frac{\gamma}{\alpha^2}.
\label{t}
\end{equation}

\begin{table}[h]
\centering
\caption{Low momentum parameters of the $\pi\pi$ scalar, $I=0$
scattering}
\vspace{.4cm}
\begin{tabular}{|c|c|c|c|}
\hline
\multicolumn{1}{|c|}{Set No} &
\multicolumn{1}{|c|}{$a_{\pi} (m_{\pi}^{-1})$} &
\multicolumn{1}{|c|}{$r_{\pi} (m_{\pi}^{-1})$} &
\multicolumn{1}{|c|}{$v_{\pi} (m_{\pi}^{-3})$} \\
\hline
\multicolumn{1}{|c|}{1} &
$0.172 \pm 0.008$ & $-8.60$ & 3.28 \\
\hline
\multicolumn{1}{|c|}{2} &
$0.174 \pm 0.008$ & $-8.51$ & 3.25 \\
\hline
\end{tabular}
\label{pi_table}
\end{table}

\vspace{.5cm}

\hspace{-.7cm}The effective range parameters are given in tables 1 and 2
for two sets of
experimental data analysed in \cite{klm}. These data sets differ
qualitatively in \\ a vicinity of the $K\overline{K}$ threshold
as shown in fig. 3 of \cite{klm}.
The $K\overline{K}$ phase shifts tend to decrease at threshold for the
set 1 and increase for the set 2.
The model \cite{klm} describes better the data set 1 than the set 2.
\begin{table}[h]
\centering
\caption{Low momentum parameters of the $K\overline{K}$ scalar, $I=0$
scattering}
\vspace{.4cm}
\begin{tabular}{|c|r|r|r|r|r|}
\hline
\multicolumn{1}{|c|}{Set} &
\multicolumn{1}{|c|}{$a_{K}$} &
\multicolumn{1}{|c|}{$r_{K}$} &
\multicolumn{1}{|c|}{$v_{K}$} &
\multicolumn{1}{|c|}{$R_{K}$} &
\multicolumn{1}{|c|}{$V_{K}$} \\
\multicolumn{1}{|c|}{No} &
\multicolumn{1}{|c|}{fm} &
\multicolumn{1}{|c|}{fm} &
\multicolumn{1}{|c|}{fm$^3$} &
\multicolumn{1}{|c|}{fm} &
\multicolumn{1}{|c|}{fm$^3$} \\
\hline
\multicolumn{1}{|c|}{1} &
$-1.73+i\,0.59$ & $-0.057+i\,0.032$ & $0.016-i\,0.0044$ & 0.38 &
-0.66 \\
\multicolumn{1}{|c|}{2} &
$-1.58+i\,0.61$ & $-0.352+i\,0.043$ & $0.028-i\,0.0057$ & 0.20 &
-0.83 \\
\hline
\end{tabular}
\label{k_table}
\end{table}

\newpage

In table 2 we have introduced two additional complex parameters
$R_{K}$ and $V_{K}$
entering into the familiar expansion valid for the real $\delta_{K}$:

\begin{equation}
k\,cot\delta_{K} = \frac{1}{Re\, a_{K}}+
\frac{1}{2} R_{K} k^2 + V_{K} k^4 +O(k^6). \label{coth}
\end{equation}
These parameters are not independent on $a_{K}, r_{K}$ and $v_{K}$ but
have been introduced for a convenience and a further discussion.
Let us notice that at least four real parametrs have to be
phenomenologically
determined in the $K\overline{K}$ channel under the condition that one
uses only two terms
of the effective range expansion (\ref{kcot}). This is in contrast to
the case
of the low energy proton--neutron scattering in the ${^3}S_{1}$ state
(as discussed by T\"ornqvist in ref. \cite{trnqv}) since in the latter
case the scattering is purely elastic.

For a full description of the two complex $\pi\pi$ and $K\overline{K}$
channels (including the
$K\overline{K} \longrightarrow \pi\pi$ annihilation process) we
introduce a real and
symmetric matrix $M$ related to the scattering matrix $T$ by
\begin{equation}
M=T^{-1}+i\,\widehat{k}
\label{m_def}
\end{equation}
where $\widehat{k}$ is a diagonal $2\times2$ matrix of the
$K\overline{K}$ and
$\pi\pi$ momenta in the \mbox{center--off--mass} system.
If we label by 1 the
$K\overline{K}$ channel and by 2 the $\pi\pi$ channel then the
$T$--matrix elements read:
\begin{equation}
T_{11}=(2ik_1)^{-1}(\eta e^{2i\delta_{1}}-1),
\label{t11}
\end{equation}
\begin{equation}
T_{22}=(2ik_2)^{-1}(\eta e^{2i\delta_{2}}-1),
\label{t22}
\end{equation}
\begin{equation}
T_{12}=T_{21}= \textstyle{1\over{2}} (k_{1}k_{2})^{-1/2} (1-\eta^2)^{1/2}
e^{i(\delta_{1}+\delta_{2})}.
\label{t12}
\end{equation}
At the $K\overline{K}$ threshold the $M$--matrix elements can be
expanded as
\begin{equation}
M_{ij}=A_{ij}+\textstyle{1\over{2}} B_{ij} k_{1}^{2}+C_{ij}k_{1}^{4}+
O(k_{1}^{6}),
\label{mij}
\end{equation}
where $A_{ij}$, $B_{ij}$ and $C_{ij}$ are real coefficients and $k_1$ is
the $K\overline{K}$ momentum (i, j=1,2). Every threshold parameter in two
channels
introduced in eq. (\ref{kcot}) can be related to a set of the $M_{ij}$
expansion parameters. For example the complex $K\overline{K}$ scattering
length is
\begin{equation}
a_{K}=\left( A_{11}-\frac{A_{12}^{2}}{A_{22}-iq}\right)^{-1},
\label{aA}
\end{equation}
\nopagebreak
where $q=(m_{K}^{2}-m_{\pi}^{2})^{1/2}$ is the pion momentum at the
$K\overline{K}$
threshold.
\newpage
We use the average pion mass
$m_{\pi}=\frac{1}{2}(m_{{\pi}^{\pm}}+m_{{\pi}^0})\approx 137.27$ MeV
and the average kaon mass
$m_{K}=\frac{1}{2}(m_{{K}^{\pm}}+m_{{K}^{0}})\approx 495.69$ MeV.
The coefficients $A_{ij}$, $B_{ij}$ and $C_{ij}$ are shown in table
(\ref{m_table}) for the data \mbox{set 1}.

\begin{table}[h]
\centering
\caption{$M$--matrix expansion parameters at the $K\overline{K}$
threshold}
\vspace{.4cm}
\begin{tabular}{|c|c|r|r|r|}
\hline
\multicolumn{1}{|c|}{reaction} &
\multicolumn{1}{|c|}{i\hspace{0.3cm}j} &
\multicolumn{1}{|c|}{$A_{ij}$} &
\multicolumn{1}{|c|}{$B_{ij}$} &
\multicolumn{1}{|c|}{$C_{ij}$} \\
\multicolumn{1}{|c|}{channel} &
\multicolumn{1}{|c|}{} &
\multicolumn{1}{|c|}{fm$^{-1}$} &
\multicolumn{1}{|c|}{fm} &
\multicolumn{1}{|c|}{fm$^3$} \\
\hline
\multicolumn{1}{|c|}{${K\overline{K}}$} &
1\hspace{.3cm}1 & -0.483 & $-8.10\times10^{-2}$ & $1.83\times10^{-2}$\\
\multicolumn{1}{|c|}{$\pi\pi$} &
2\hspace{.3cm}2 & 0.476 & $-1.58\times10^{-1}$ & $1.43\times10^{-3}$ \\
\multicolumn{1}{|c|}{$K\overline{K} \longleftrightarrow \pi\pi$} &
1\hspace{.3cm}2 & 0.669 & $-1.57\times10^{-2}$ & $5.93\times10^{-3}$ \\
\hline
\end{tabular}
\label{m_table}
\end{table}

\vspace{.6cm}

At first let us discuss the $\pi\pi$ threshold parameters.
The $\pi\pi$ scattering length is small and positive while the
$\pi\pi$ effective range is negative and much larger.
The third parameter (sometimes called the shape
parameter) is positive in our model. In a recent analysis of the near
threshold $\pi N \longrightarrow \pi\pi N$ data D. Po\v{c}ani\'c et al.
\cite{pocanic} have provided the $\pi\pi$ scattering length
\mbox{$a= (0.177 \pm 0.006)$} \mbox{$m_{\pi}^{-1}$}
which is in a very good agreement with our predictions \cite{klm}
(compare the second column of table 1).
In the earlier analyses Lowe et al. \cite{lowe} and
Burkhardt and Lowe \cite{burkhardt} have given the $\pi\pi$ scattering
length values
$(0.207 \pm 0.028)\; m_{\pi}^{-1}$ and $(0.197 \pm 0.01)\;
m_{\pi}^{-1}$, respectively.
Using the chiral perturbation theory Gasser and Leutwyler
\cite{gasser} have
obtained a value $(0.20 \pm 0.01)\; m_{\pi}^{-1}$ while in a recent
paper by Roberts
et al. \cite{roberts} the calculated values of the scattering length
are 0.16 $m_{\pi}^{-1}$ or 0.17 $m_{\pi}^{-1}$.

The $\pi\pi$ effective range is not well determined experimentally.
Belkov et al.
\cite{belkov79} have obtained $r_{\pi} =(-9.6 \pm 19.1)\;
m_{\pi}^{-1}$. Based
on the analysis of the $\pi^{-} p \longrightarrow \pi^{+}\pi^{-} n$
data
performed by Belkov and Buniatov \cite{belkov82} we have derived the
value of
the effective range $r_{\pi}=-8.1\; m_{\pi}^{-1}$ with an estimated
error at least 65\%. Within the Weinberg
approach \cite{weinberg66} the parameter
$r_{\pi} = -8.48\; m_{\pi}^{-1}$
which is very close to our values about
$-8.6\; m_{\pi}^{-1}$ or $-8.5\;
m_{\pi}^{-1}$ given in table 1 (the scattering length used in the
Weinberg model was 0.157 $m_{\pi}^{-1}$).
The effective range $(-7.4 \pm 2.5)\;
m_{\pi}^{-1}$ can be obtained from two low energy parameters
$a$ and $b$
predicted in ref. \cite{gasser}. It is also possible to evaluate the
effective
range from the similar parameters fitted to the $\pi\pi$  phase shifts
by Rosselet
et al. \cite{rosselet} in the study of the $K_{e4}$ decays
\mbox{($a=0.28 \pm 0.05$}, \mbox{$ b=0.19-(a-0.15)^2$)}.
Its value is ($-1.4 \pm 3.7$) $m_{\pi}^{-1}$ which is
considerably different from the above cited value $-8.5$ fm.
Another estimation
based on the same data using $a$ and $b$ as free parameters leads to
a different
value $r_{\pi}=(0.3 \pm 6.3)\; m_{\pi}^{-1}$. We infer from these
numbers that the
existing $\pi\pi$ data are not yet substantially accurate to determine
the effective range with a good precision.

The effective range expansion (\ref{kcot}) in the $\pi\pi$ channel
has a limited convergence range due to
a presence of the left--hand cuts in the Mandelstam variable
$s=4(m_{\pi}^2+k^2)$. In the momentum plane $k$ there are two cuts
starting at
$k=\pm i m_{\pi}$ (see also fig. 5 of ref. \cite{klm}). These cuts lie
very
close to the $\pi\pi$ threshold and lead to a negative contribution
to the
$\pi\pi$ scattering length ($-0.18\; m_{\pi}^{-1}$). The second
negative contribution
($-0.24\; m_{\pi}^{-1}$) comes from the
singularities of  the $\pi\pi$ interaction. The dominant positive
contribution to
$a_{\pi}$ has its origin in a presence of the $f_{0}(500)$ pole in the
$\pi\pi$ scattering
amplitude ($+0.60\; m_{\pi}^{-1}$). In the practical applications of
the effective
range formula the experimental data should be carefuly selected from a
$\pi\pi$
momentum range very close to the threshold in order to diminish the
contribution
of higher terms usually neglected in the analyses.
The $\pi\pi$ energy corresponding
to the maximum momentum at which the convergence limit is attained in
the presence of the above--mentioned cuts is as low as 390 MeV.

The $K\overline{K}$ scattering length is complex in presence of the
open annihilation
channel. Modulus of its real part is much larger than the $\pi\pi$
scattering length. The
imaginary part is positive and gets a value about 0.6 fm. As seen in
\mbox{table \ref{k_table}}
the expansion parameters $r$ and $v$ are rather small. This is not
accidental
and can be easily understood if one notices a fact that the
$S$--matrix pole $f_{0}(975)$
is very close to the $K\overline{K}$ threshold. Its position in the
$K\overline{K}$ momentum frame is
$p_{0}=(-34.7+i\, 100.3)$ MeV for the set 1 and
$p_{0}=(-36.1+i\, 100.2)$ MeV for the set 2.
If we approximate the $K\overline{K}$ element of the $S$--matrix by
its dominant pole contribution:

\begin{equation}
S_{K\overline{K}}^{^{pole}}=\frac{-k-p_{0}}{k-p_{0}},
\label{spole}
\end{equation}
then the $K\overline{K}$ scattering length is $a_{0}=(i p_{0})^{-1}$
(see also ref.
\cite{marsylia}) and all other parameters of the threshold expansion of
$k\,cot \delta$ identically vanish since $k\,cot \delta \equiv 1/a_{0}$.
Therefore in the single $f_{0}(975)$ pole approximation the parameters
$r_{K}$ and $v_{K}$
are zero. Their smalleness in the full model  calculation is a reflection
of the
$f_{0}(975)$ dominance near the $K\overline{K}$ threshold.
The values $a_{0}$ are ($-1.76 + i\, 0.61$) fm
for the set 1 and ($-1.74 + i\, 0.63$) fm for the set 2; they are quite
close to the
values $a_{K}$ given in table 2 especially for the set 1 preffered by
our model.
The negative sign of Re$a_{K}$ is characteristic for the appearence of
a bound
$K\overline{K}$ state $f_{0}(975)$.
We have studied an accuracy of the pole approximation (\ref{spole})
in comparison
with the results calculated from the complete model. For the model
parameters
fitted to the data set 1 both the $K\overline{K}$ phase shifts and
the inelasticity are
reproduced with a precision better than 2\% for the $K\overline{K}$
momenta as large as 380
MeV/c (or the effective mass as high as 1250 MeV). For the set 2 the
inelasticity
parameter is described within 3\% up to 450 MeV/c but the phase shifts
are less
accurately reproduced (to 11\% at the threshold and up to 17\%
at 400 MeV/c). At the energies higher than
1250 MeV the $f_{0}(1400)$ resonance plays an important role and gives
an additional contribution to the $f_{0}(975)$ term.

The $K\overline{K}$ effective range parameter $R_{K}$ is relatively
small in comparison with $\mid$ Re $a_{K} \mid$.
The contribution of the $f_{0}(975)$ pole to the third parameter $V_{K}$
shown in table 2
is also dominant.
In this approximation both parameters $R_{K}$ and $V_{K}$ are given in
terms of Re$a_{K}$ and Im$a_{K}$.
If the kaon momentum increases then the higher terms in the
threshold expansion become important.
The convergence radius of the expansions (\ref{delta}) and (\ref{coth})
is equal to a distance $\mid p_{0} \mid$ to the nearest
$S$--matrix pole.
The energy
corresponding to $k=\mid p_{0} \mid$ is 1014 MeV which is only 23 MeV
above the
$K\overline{K}$ threshold.
Therefore one can draw a severe limit on the experimental energy
resolution needed in
the determination of the $K\overline{K}$ threshold parameters.
In practice one should require
the energy resolution of the order of 1 MeV.
The expansion (\ref{mij}) of the $M$--matrix, however,
has a larger convergence radius
495.69 MeV/c limited by the kaon mass.

According to our knowledge the experimental information about the
$K\overline{K}$ threshold
parameters is almost nonexistent. We are aware of only one pioneer
experimental
determination of the $K_{S}^{0}K_{S}^{0}$ scattering length by
Wetzel et al. \cite{wetzel}.
Although the values obtained by authors of \cite{wetzel}
($\mid a \mid = (1.25 \pm 0.12)$
fm, Im$a = (0.27 \pm 0.03)$ fm) are of the same order as our
determinations, we
think that their errors are too small. There are at least two reasons
to believe
that this observation is true: firstly only two experimental points are
used in
the analysis for the $K\overline{K}$ effective mass smaller than
1.1 GeV and secondly their
parametrization of the $K\overline{K}$ phase shifts does not fulfil
the general symmetry
requirement: $\delta_{K\overline{K}}(-k)=\delta_{K\overline{K}}(k)$.
Nevertheless these data seem to
indicate a fact that the modulus of the $K\overline{K}$ scattering
length is much larger than the $\pi\pi$ scattering one.

In conclusion, we have determined the effective range parameters of the
$\pi\pi$ and $K\overline{K}$ scalar--isoscalar interactions. We hope
that our predictions will be confronted in future with new data clearly
needed to understand the nature of the scalar mesons.

\vspace{.5cm}

This work has been partially supported by the Polish Committee for
Scientific Research (grant No 2 0198 9101) and by Maria
\mbox{Sk\l{}odowska--Curie} Fund II (No PAA/NSF--94--158).
We thank very much J.-P. Maillet for the discussions.

\end{document}